\documentclass[lettersize,journal]{IEEEtran}
\usepackage{amsmath,amsfonts}
\usepackage{algorithmic}
\usepackage{algorithm}
\usepackage{array}
\usepackage[caption=false,font=normalsize,labelfont=sf,textfont=sf]{subfig}
\usepackage{textcomp}
\usepackage{stfloats}
\usepackage{url}
\usepackage{verbatim}
\usepackage{graphicx}
\usepackage{siunitx}
\usepackage{balance}
\usepackage{cite}

\title{Lossy Compression of Network Feature Data: When Less Is Enough}

\author{
    \IEEEauthorblockN{
        Fabio Palmese\IEEEauthorrefmark{2},
        Gabriele Merlach\IEEEauthorrefmark{1},
        Damiano Ravalico\IEEEauthorrefmark{1},
        Martino Trevisan\IEEEauthorrefmark{1},
        Alessandro E. C. Redondi\IEEEauthorrefmark{2}
    }
    
    \IEEEauthorblockA{
        \IEEEauthorrefmark{1}University of Trieste, 
        \IEEEauthorrefmark{2}Politecnico di Milano
    }
    \thanks{F. Palmese  and A. E. C. Redondi, are with, Politecnico di Milano, Milan, Italy. G. Merlach, D. Ravalico and M. Trevisan are with the University of Trieste. This study was carried out within the PRIN project COMPACT and received funding from Next Generation EU, Mission 4 Component 1, CUP: D53D23001340006.}
}
\begin{document}

\maketitle

\begin{abstract}
Network traffic analysis increasingly relies on feature-based representations to support monitoring and security in the presence of pervasive encryption. 
Although features are more compact than raw packet traces, their storage has become a scalability bottleneck from large-scale core networks to resource-constrained Internet of Things (IoT) environments. 
This article investigates task-aware \emph{lossy} compression strategies that reduce the storage footprint of traffic features while preserving analytics accuracy. 
Using website classification in core networks and device identification in IoT environments as representative use cases, we show that simple, semantics-preserving compression techniques expose stable operating regions that balance storage efficiency and task performance. 
These results highlight compression as a first-class design dimension in scalable network monitoring systems.
\end{abstract}

\begin{IEEEkeywords}
Network traffic analysis, feature-based monitoring, lossy compression, network analytics.
\end{IEEEkeywords}

\section{Introduction}
\label{sec:intro}

Network traffic analysis is an essential capability for modern communication networks, underpinning a wide range of operational functions including performance monitoring, capacity planning, security enforcement, and digital forensics. Over the last decade, however, the operational assumptions that traditionally guided traffic analysis systems have been progressively eroded. Two trends in particular are reshaping how network data can be collected, stored, and exploited.

First, network traffic volumes and rates have grown dramatically, driven by bandwidth-intensive services and by the massive proliferation of connected devices \cite{cisco2020internet}. Even short observation windows in backbone, access, or edge networks can generate enormous amounts of measurement data. Second, the widespread adoption of encryption at the transport and application layers has rendered classical packet inspection techniques largely ineffective \cite{naylor2014cost}. As a consequence, modern traffic analysis pipelines no longer operate using the so-called Deep Packet Inspection'' (DPI), but instead rely on statistical descriptors---traffic features---that capture timing, size, and behavioral characteristics of flows or devices \cite{shen2022machine}.

This shift toward feature-based traffic analysis has enabled accurate monitoring and inference even in fully encrypted environments and has accelerated the adoption of machine learning techniques across network operations and security. At the same time, it has introduced a less visible but increasingly critical bottleneck: the storage and long-term management of traffic features themselves. While feature representations are substantially more compact than raw packet traces, their volume remains challenging in realistic deployments. In large core networks, feature datasets can easily reach tens of gigabytes per day even after standard lossless compression \cite{trevisan2018five}, while in Internet of Things (IoT) environments, storage constraints at gateways and access points are often severe, especially when long-term evidence retention is required.

In practice, traffic features are derived using different aggregation strategies depending on the scenario---for example, flow-level summaries in core networks or window-based device activity descriptors in IoT settings---but they ultimately serve the same purpose: acting as the input to downstream analytics pipelines whose accuracy must be preserved. Current storage practices typically rely on generic data formats and compression mechanisms that are agnostic to both the semantics of traffic features and the requirements of the analysis tasks they support \cite{hofstede2014flow}.

\begin{figure*}[t]
    \centering
    \includegraphics[width=0.8\textwidth]{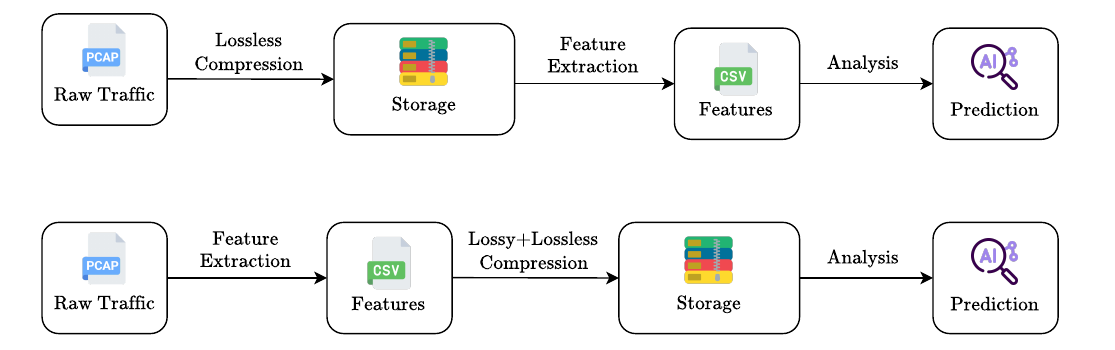}
    \caption{Conventional network traffic analysis pipelines typically apply compression only to raw packet traces, while extracted network feature data are stored at full numerical precision (top). This work examines an alternative design in which compression---potentially including lossy steps---is applied directly to network feature data, reducing storage requirements while preserving the utility of downstream analytics (bottom).}
    \label{fig:pipeline}
\end{figure*}

Figure~\ref{fig:pipeline} illustrates this shift in perspective. Conventional traffic analysis pipelines (first row) apply compression primarily to raw packet traces, while extracted feature data are typically stored at full numerical precision. In contrast, we consider an alternative design in which compression---potentially including lossy steps---is applied directly in the feature space (second row), reflecting the fact that the ultimate value of these data lies in their support of downstream analytics rather than exact numerical reconstruction.

In this work, we explore lossy compression strategies for network feature data as a principled way to address this tension. Rather than treating features as signals that must be preserved with full numerical fidelity, we adopt a task-aware perspective and evaluate compression techniques directly in terms of their impact on downstream analytics accuracy. This perspective reflects how modern monitoring systems are used in practice: the utility of stored data is ultimately determined by task performance, not by reconstruction precision.

We ground our analysis in two complementary and practically relevant scenarios spanning modern network architectures. The first considers core network traffic analysis, where features extracted at scale are used to infer contacted services under encryption. The second focuses on IoT device identification in smart home environments, where gateways and access points operate under severe resource constraints while supporting long-term monitoring and forensic needs.
By grounding compression decisions directly in task performance, this work provides actionable insights for designing scalable, feature-based network monitoring systems across both centralized core networks and resource-constrained IoT environments.

The remainder of this article is organized as follows. We first examine feature compression in large-scale core network traffic analysis, and then focus on IoT device identification in resource-constrained environments. We conclude by distilling cross-scenario design guidelines and outlining open challenges for scalable, feature-based network monitoring.

\section{Background and State of the Art}

The problem of collecting, storing, and analyzing network traffic data has been extensively studied over the past two decades, leading to a rich body of work spanning traffic compression, flow-level monitoring, and feature-based analytics. 
However, much of this literature is grounded in assumptions that are increasingly misaligned with the landscape of modern networks, particularly in the presence of pervasive encryption and large-scale, long-term monitoring requirements.

\subsection{Packet-Level Traffic Compression}

Early work on network traffic compression focused on reducing the storage footprint of packet traces, motivated by the high cost of retaining full packet captures for offline analysis~\cite{holanda2005performance,chen2008ipzip}. 
These approaches exploit redundancy in packet headers, payloads, and protocol semantics, and are effective when packet contents are available and compression is applied close to the data source.

The widespread adoption of encryption at the transport and application layers has fundamentally limited the applicability of packet-level compression. 
Encrypted payloads largely eliminate content redundancy, while header compression alone provides limited gains. 
As a result, packet-centric approaches no longer offer a viable solution for scalable traffic storage in operational networks.

\subsection{Flow-Level Monitoring and Feature-Based Representations}

To address scalability and privacy concerns, network operators have progressively shifted from packet-level monitoring to flow-level and feature-based representations of traffic, as exemplified by NetFlow and IPFIX~\cite{hofstede2014flow}. 
Flow records and derived features provide a compact summary of network activity while enabling a wide range of analytics tasks, including traffic classification, performance monitoring, and security analysis.

Despite their reduced size compared to packet traces, flow records and feature logs can still generate substantial data volumes in large-scale deployments~\cite{trevisan2018five}. 
Existing solutions typically rely on generic lossless compression or database-level optimizations, implicitly assuming that feature values must be preserved with full numerical precision. 
This assumption is rarely questioned, even though downstream analytics often operate on statistical patterns rather than exact values.

\subsection{Dimensionality Reduction and Learning-Based Approaches}

A parallel line of research has explored the use of dimensionality reduction and learning-based techniques to compact traffic representations. 
Methods such as Principal Component Analysis (PCA), clustering, and neural autoencoders aim to project high-dimensional feature spaces into lower-dimensional embeddings while preserving information content~\cite{shen2022machine,pekar2025autoflow}.

These approaches are typically evaluated in terms of reconstruction error or variance preservation, implicitly assuming that statistical fidelity correlates with analytical utility. 
However, such metrics are poorly aligned with storage efficiency and downstream task performance. 
Moreover, transformed or learned representations often increase data entropy and introduce additional computational complexity, which can undermine practical compressibility and scalability in high-throughput monitoring pipelines.

\subsection{IoT Traffic Monitoring and Forensic Pipelines}

In Internet of Things (IoT) environments, traffic monitoring has been widely studied in the context of device identification, activity inference, and forensic investigations. 
Many proposed pipelines rely on collecting packet traces at gateways and offloading data to centralized servers or cloud infrastructures for analysis~\cite{sivanathan2018classifying}. 
While effective in controlled settings, these approaches raise significant concerns related to storage cost, data privacy, regulatory compliance, and dependence on continuous connectivity.

Feature-based monitoring has emerged as a promising alternative, enabling privacy-preserving analytics directly at the network edge~\cite{palmese2023designing}. 
However, existing IoT monitoring solutions often assume either short retention periods or ample backend storage, and rarely address the problem of long-term feature storage under strict resource constraints.

\subsection{Open Challenges}

From core networks to IoT environments, existing work lacks a unified view of compression that explicitly accounts for the requirements of downstream analytics.
In particular, there is limited understanding of how lossy compression of traffic features impacts task accuracy, how different compression levers behave across architectures, and how storage efficiency should be measured in a deployment-aware manner.

This article addresses these gaps by adopting a task-aware perspective on feature compression and by evaluating its impact across two representative and complementary scenarios. 
Rather than proposing new compression algorithms, we focus on extracting practical design lessons that can guide the deployment of scalable, analytics-driven monitoring systems in modern networks.

\section{Core Network Traffic Analysis}
\label{sec:core}

Large-scale core networks represent one of the most demanding environments for traffic analytics. 
Internet Service Providers and backbone operators must continuously monitor traffic generated by large populations of users at multi-Gbit/s rates, while retaining sufficient information to support performance monitoring, traffic characterization, and security analysis. 
In these settings, feature-based representations of traffic are essential: packet payloads are typically encrypted, and analytics pipelines operate on statistical descriptors extracted from observed traffic.

\subsection{Use Case: Domain Classification}

We consider \emph{domain classification} as a representative analytics task for core network monitoring. 
The objective is to infer the server domain contacted by a client from encrypted traffic, a problem that reflects practical ISP needs for service-level monitoring without payload inspection and has been widely used to assess the limits of traffic analysis under encryption.

Our evaluation relies on a large flow-level dataset collected over a full day from a university campus network. 
Traffic was captured by a passive probe deployed at the campus edge router and aggregated into flow records using the \emph{Tstat} traffic analyzer. 
Each flow record summarizes a TCP or UDP connection through a rich set of statistical descriptors capturing packet sizes, timing behavior, retransmissions, and flow duration, as well as metadata such as domain names extracted from the TLS Server Name Indications (SNIs) field in the Client Hello messages. 
In total, more than \num{175} numerical features are extracted per flow and constitute the feature space targeted by compression, while non-numerical fields are excluded from lossy processing. 
IP addresses are anonymized and sensitive information is discarded to preserve user privacy.

Over a single day of observation, the dataset includes more than \num{14} million flows generated by several thousand client IP addresses communicating with hundreds of thousands of servers. 
To ensure robustness and operational relevance, we focus on domains associated with at least \num{1000} observed flows and hosted by prominent Autonomous Systems (ASes). 
This selection results in \num{35} ASes, including major content providers, CDNs and cloud service platforms, which collectively account for approximately \qty{90}{\percent} of the observed traffic. 
Domains are labeled using HTTPS SNIs, while less frequent domains are grouped into a single \emph{Other} class, yielding an intentionally imbalanced classification problem. 
Overall, the task involves \num{708} distinct domain classes.

To manage this scale and reflect practical deployments, we adopt an AS-based modeling strategy: a separate classifier is trained for each AS, and classification is restricted to flows associated with that AS. 
This design mirrors operational pipelines, where the AS of a server can be inferred reliably from network-layer information and significantly reduces the number of candidate domains considered by each model. 
Classification is performed using Random Forest models, chosen for their robustness and established effectiveness in encrypted traffic analysis. 
Performance is quantified using the F1-score, which captures the balance between precision and recall and is well suited to the inherent class imbalance of the task.

\subsection{Storage Footprint and Compression Strategies}

Although feature extraction dramatically reduces the volume of traffic data compared to raw packet traces, the resulting storage footprint remains operationally significant at backbone scale. When feature logs are stored using standard lossless compression (e.g., \texttt{gzip}), the daily footprint is approximately \qty{14.9}{\giga\byte}---an \qty{84}{\percent} reduction relative to raw packet captures. While this reduction is substantial, such volumes still accumulate rapidly under long-term retention policies, motivating additional compression applied directly to the feature representation.

To this end, we investigate lossy compression in the numerical feature space through feature-wise uniform scalar quantization. Each numerical feature is discretized independently by mapping its continuous values onto a finite number of uniformly spaced levels determined by a chosen bit width. The number of quantization bits controls the granularity of representation: lower bit depths reduce storage but introduce stronger numerical coarsening.

To ensure methodological rigor and prevent information leakage, quantization ranges are defined per feature using only the training data. Specifically, for each feature we estimate its minimum and maximum values from the training split and use this fixed range to discretize both training and test samples. Quantization is therefore performed after the train/test split, with no statistics derived from the test set influencing the encoding process. We do not recompute ranges across time windows; instead, a single global range per feature (derived from training data) is applied consistently throughout the evaluation. This design guarantees reproducibility and reflects how a deployed monitoring system would operate once configured.

Following quantization, the resulting feature logs are compressed using standard lossless tools (e.g., \texttt{gzip}). Reported storage reductions thus capture the end-to-end footprint, combining the effect of controlled numerical precision reduction with practical lossless compression, as would occur in operational telemetry pipelines.

We also investigate the interaction between scalar quantization and dimensionality reduction based on Principal Component Analysis (PCA). 
PCA projects features into a lower-dimensional space while preserving a target fraction of statistical variance. 
We evaluate two PCA operating points:

\begin{itemize}
    \item Gentle PCA: preserves \qty{99}{\percent} of total variance.
    \item Aggressive PCA: preserving \qty{80}{\percent} of total variance.
\end{itemize}

Both are combined with scalar quantization, to assess whether variance-preserving projections translate into improved compressibility and task performance.

\subsection{Accuracy--Storage Tradeoffs}

\begin{figure}
    \centering
    \includegraphics[width=\columnwidth]{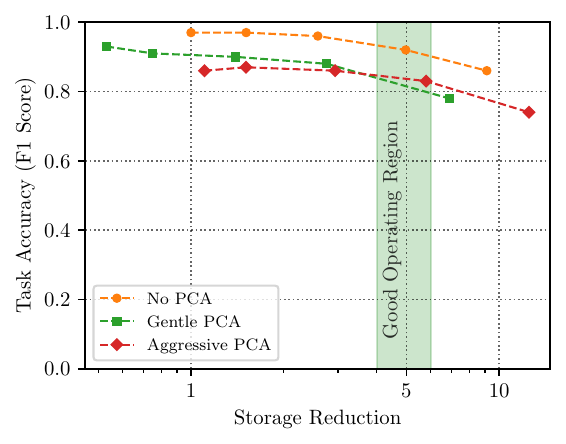}
\caption{Accuracy--storage tradeoff for core network domain classification. 
Feature-wise scalar quantization achieves substantial additional storage reduction with limited accuracy loss, while PCA-based compression leads to inferior operating points. 
The shaded area indicates a practical operating region ($4\text{--}6\times$ storage reduction). From left to right, the markers refer to quantizing each feature with 32, 16, 8, 4 and 2 bits.}
    \label{fig:core}
\end{figure}

Figure~\ref{fig:core} reports the tradeoff between \emph{storage reduction} and \emph{task accuracy} for the core network domain classification task. 
The vertical axis shows the F1-score, while the horizontal axis reports the achieved storage reduction factor, measured relative to storing feature logs compressed with lossless compression alone.

The curve labeled \emph{Scalar Quantization (No PCA)} corresponds to feature-wise scalar quantization, where each marker corresponds to a different bit depth. 
The point at storage reduction equal to one represents the baseline case in which features are stored without lossy quantization and only lossless compression is applied, yielding an F1-score of approximately \num{0.97}. 
As quantization becomes more aggressive, numerical precision is progressively reduced and additional storage savings are obtained. 
Accuracy degrades gradually: even when achieving nearly an order-of-magnitude additional reduction over lossless-compressed feature logs, the F1-score decreases by less than \num{0.1}. 
This behavior highlights the robustness of feature-based analytics to quantization noise.

The shaded area in Figure~\ref{fig:core} identifies a \emph{practical operating region}, corresponding to an additional $4\text{--}6\times$ storage reduction over lossless-compressed feature logs. 
In this region, accuracy remains close to the non-quantized baseline while providing a meaningful multiplicative reduction on top of the already significant \qty{84}{\percent} savings achieved through feature extraction and lossless compression. 
From an operational perspective, this region represents a stable and tunable compromise between storage efficiency and analytics accuracy.

In contrast, the curves corresponding to scalar quantization combined with PCA exhibit a markedly different behavior. 
While gentle PCA preserves accuracy at low compression levels, more aggressive quantization rapidly degrades task performance. 
Aggressive PCA further amplifies this effect. 
Projecting features into a transformed space alters their statistical structure, often increasing data entropy and reducing the effectiveness of subsequent lossless compression. 
As a result, PCA-based dimensionality reduction may fail to deliver the expected storage savings and leads to inferior operating points compared to direct feature-wise quantization.

We also evaluated vector quantization approaches, where groups of features are jointly encoded by clustering feature vectors and representing them through cluster centroids. 
While such techniques can, in principle, achieve high compression ratios, we found them to be poorly suited for large-scale core network settings. 
The computational and memory overhead associated with clustering high-dimensional feature spaces, together with limited robustness under traffic heterogeneity, makes vector quantization difficult to deploy in high-throughput monitoring pipelines. 
As a result, vector-based approaches do not provide a favorable accuracy--storage--complexity tradeoff compared to simple feature-wise quantization in this scenario \cite{merlach2025handling}.

\subsection{Implications for Core Network Monitoring}

These results convey a clear design lesson for core network monitoring systems. 
Additional storage gains beyond lossless compression are feasible and operationally meaningful, even if the absolute compression factors may appear modest when considered in isolation. 
Importantly, these gains are achieved on top of already highly compressed feature logs, without modifying existing monitoring pipelines or storage backends.

At the same time, preserving statistical variance through dimensionality reduction does not guarantee improved compressibility or analytical robustness. 
In large-scale core networks, simple and semantics-preserving compression techniques such as scalar quantization provide a more reliable and deployable solution than more sophisticated transformations.

In the next section, we show how a similar task-aware compression perspective applies to IoT device identification, where different constraints and compression levers lead to markedly different operating points.

\section{IoT Traffic Analysis}

Internet of Things (IoT) environments pose a complementary set of challenges for traffic analytics compared to core networks. 
While traffic volumes are typically consistently lower, IoT gateways and WiFi access points often operate under strict constraints in terms of storage capacity, processing power, and long-term data retention.
Although cloud-based storage can be used to offload data, it introduces additional challenges related to operational costs, data privacy, regulatory compliance, and dependence on continuous connectivity.
At the same time, IoT deployments increasingly require continuous monitoring to support security, troubleshooting, and forensic investigations. 
In these settings, feature-based representations of traffic are again essential, as payloads are encrypted and analytics must rely on statistical descriptors of device behavior.

\subsection{Use Case: IoT Device Identification}

We consider \emph{IoT device identification} as a representative analytics task for smart home and edge environments. 
The objective is to identify the specific device model responsible for observed network traffic, solely based on traffic features. 
Accurate device identification is a fundamental building block for a wide range of applications, including security monitoring, anomaly detection, and forensic analysis. 
Moreover, identifying the device type enables selective monitoring strategies, such as activating task-specific analytics pipelines only for certain devices (e.g., cameras or voice assistants) or filtering out non-IoT traffic to reduce storage requirements.

Our evaluation relies on a publicly available dataset containing network traffic traces from a diverse set of consumer IoT devices operating in a smart home environment \cite{sivanathan2018classifying}. 
Traffic is captured at the network gateway and processed to extract feature-based representations that summarize device activity over fixed time windows in the order of 5-10 seconds. 
Each device's traffic is summarized through a set of \num{140} statistical features capturing packet sizes, payload sizes, interarrival times, packet length distributions, and basic addressing information. 
To ensure realistic operating conditions, we consider a heterogeneous mix of IoT devices, including sensors, actuators, and multimedia devices, as well as a limited number of non-IoT devices. 
The resulting classification task involves multiple device classes with uneven traffic volumes, reflecting the natural imbalance observed in real deployments. 
Device identification is formulated as a supervised learning problem, and performance is evaluated using the F1-score, which captures the balance between precision and recall across device classes.

\subsection{Storage Footprint and Feature Selection}

Unlike core networks, where feature volumes are dominated by traffic scale, storage requirements in IoT environments are driven by the need for \emph{long-term retention} under limited local resources. 
Even moderate per-device data rates can accumulate into substantial storage footprints over weeks or months of continuous monitoring.

To control data growth, we leverage two complementary compression levers.
First, we apply \emph{feature selection} to reduce the dimensionality of the feature space by retaining only the most informative features for the device identification task. 
This step directly reduces the number of values that must be stored per time window and is particularly effective in IoT settings, where many extracted features may be redundant or weakly informative. We consider:
\begin{itemize}
\item Moderate selection: keeping the most informative 100 features.
\item Aggressive selection: considering only 20 features among the original ones. 
\end{itemize}

Second, as in the core network scenario, we apply \emph{feature-wise scalar quantization} to the selected features, reducing numerical precision while preserving feature semantics. 
After quantization, feature logs are stored using standard lossless compression (e.g., gzip), reflecting practical deployment conditions. 
Storage efficiency is therefore evaluated in terms of the end-to-end storage rate required to support the analytics task.

\subsection{Accuracy--Storage Tradeoffs}

\begin{figure}
    \centering
    \includegraphics[width=\columnwidth]{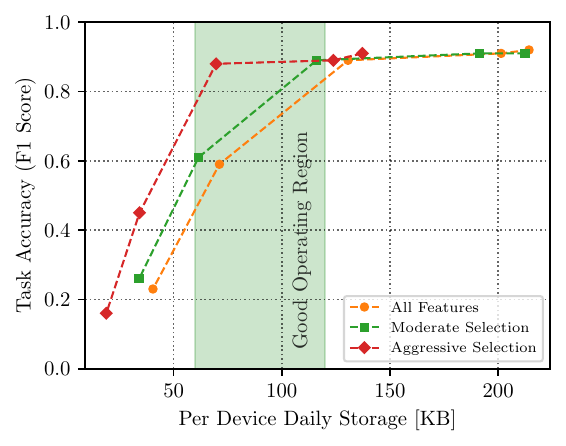}
    \caption{Accuracy--storage tradeoff for IoT device identification. 
Feature selection combined with scalar quantization enables near-maximal accuracy at very low storage rates per device; the shaded area highlights a practical operating region suitable for long-term retention on resource-constrained IoT gateways. From left to right, the markers refer to quantizing each feature with 2,4,8,16 and 32 bits.}
    \label{fig:iot}
\end{figure}

Figure~\ref{fig:iot} illustrates the tradeoff between task accuracy and storage rate for the IoT device identification task under different levels of feature selection and quantization. 
The vertical axis reports the F1-score, while the horizontal axis shows the required storage rate, expressed as the number of bits per second needed to store feature logs.

The baseline configuration (orange line) corresponds to using the full feature set without lossy quantization, yielding the highest classification accuracy. 
In our dataset, this baseline already reflects a substantial reduction with respect to raw traffic: over $14$ days of monitoring $20$ devices, lossless-compressed packet traces amount to approximately \qty{13}{\mega\byte} per device per day, while lossless-compressed feature logs require only about \qty{210}{\kilo\byte} per device per day, corresponding to roughly $20$ bit/s per device. 
These values define the starting operating point for the IoT scenario.

As the number of selected features is reduced, storage requirements decrease significantly, with only a moderate impact on accuracy. 
Applying scalar quantization further reduces the storage rate, and accuracy degrades gradually as numerical precision is reduced.

The results reveal a wide operating region in which accurate device identification can be achieved with very low storage requirements. 
In particular, near-optimal F1-scores are maintained at storage rates on the order of a few hundred bits per second per device. 
At these rates, the daily storage footprint remains on the order of a few megabytes per device, making continuous, long-term monitoring feasible even on resource-constrained gateways.

Compared to the core network scenario, IoT environments exhibit a more favorable accuracy--storage tradeoff. 
This difference stems from the stronger structure and regularity of IoT device traffic, which allows device-specific behavioral patterns to be captured with a compact set of features and coarse numerical precision.

\subsection{Implications for IoT Monitoring}

These results highlight that, in IoT environments, \emph{feature selection} is a powerful complement to quantization for controlling storage requirements. 
By jointly selecting informative features and reducing numerical precision, it is possible to achieve substantial additional reductions in storage requirements while preserving reliable IoT device identification.

From a system design perspective, this enables practical deployment of continuous IoT monitoring and forensic-ready gateways without requiring large storage capacities or centralized data offloading. 
At the same time, the task-aware compression perspective remains essential: aggressive reduction beyond the stable operating region leads to abrupt accuracy degradation, underscoring the need to explicitly balance storage efficiency and analytics performance.

Together with the core network results, these findings demonstrate that while different environments expose different compression levers, a common principle applies across architectures: compression strategies should be evaluated and tuned based on their impact on task accuracy, rather than on abstract notions of signal fidelity alone.

\section{Design Lessons and Open Challenges}

\begin{table}[t]
\centering
\caption{Storage footprint at different stages of the monitoring pipeline for the two use cases. Values are normalized per day.}
\label{tab:storage_summary}
\begin{tabular}{lcc}
\hline
\textbf{Data Representation} & \textbf{Core Network} & \textbf{IoT} \\
\hline
Raw traces & $\approx$\qty{93}{\giga\byte} / day & $\approx$\qty{29}{\mega\byte} / device \\
Extracted features (lossless) & $\approx$\qty{14.9}{\giga\byte} / day  & $\approx$\qty{210}{\kilo\byte} / device \\
Lossy compressed features & $\approx$$3$--\qty{5}{\giga\byte} / day  & $\approx$\qty{70}{\kilo\byte} / device \\
\hline
\end{tabular}
\end{table}

Table~\ref{tab:storage_summary} provides a consolidated view of the storage footprint at different stages of the monitoring pipeline for the two use cases considered in this article, from raw traffic to task-aware compressed feature logs. The core traffic and IoT case studies sit at opposite ends of the network spectrum: high-throughput backbone monitoring on one side, and resource-constrained gateways on the other. Yet both expose the same fundamental question: how much precision do we really need to store in order to keep analytics useful?

Two complementary effects emerge from the table. In the core network scenario, feature extraction and lossless compression already achieve the dominant reduction in storage volume, while task-aware compression provides additional, operationally meaningful savings on top of an optimized baseline. In contrast, in the IoT scenario, feature extraction dramatically reduces absolute storage requirements, and task-aware compression further lowers per-device storage from hundreds to tens of kilobytes per day, enabling long-term retention on commodity gateways.

These observations motivate a set of practical design guidelines and highlight several open challenges for scalable, feature-based network monitoring systems.

\subsection{Key Design Guidelines}

\textbf{1) Compress features, not packets.} 
Once encryption becomes the norm, packet payloads stop being a useful compression target, while feature-based analytics remain viable. 
Compression effort should therefore focus on the numerical feature space, where task relevance is explicit, and precision can be traded for storage.

\textbf{2) Measure compression by its impact on the task.} 
In both scenarios, sizable reductions in storage are possible with limited loss in F1-score, even when feature values are heavily quantized or selected. 
This confirms that generic distortion metrics or variance preservation are poor proxies for utility. 
Compression should be tuned and evaluated directly against the performance of the target analytics task.

\textbf{3) Favor simple, semantics-preserving transformations.} 
In core networks, feature-wise scalar quantization provides robust, easily deployable savings on top of lossless compression, while more complex projections such as PCA increase complexity and often lead to worse accuracy--storage operating points. 
In IoT settings, feature selection plays a similar role: eliminating weakly informative dimensions before quantization is more effective than compressing a large, redundant feature set.

\textbf{4) Think in terms of operating regions, not single optima.} 
Neither scenario admits a unique ``best'' compression configuration. 
Instead, both expose stable regions where storage can be tuned without abruptly degrading accuracy. 
Systems should be designed to make these tradeoffs explicit, so that operators can move along the curve as storage, accuracy, or regulatory requirements evolve.

\subsection{Open Challenges}

\textbf{Adaptive compression.} 
Our study considers static configurations; in practice, traffic intensity, application mix and importance of specific tasks change over time. 
Designing mechanisms that adjust feature selection and quantization online, based on observed traffic and accuracy feedback, remains an open problem.

\textbf{Pushing compression into the network.} 
Programmable switches and smart gateways make it possible to perform feature extraction, compression, and even inference in-network or at the edge. 
Realizing this vision requires high-speed packet capture and processing, lightweight models, predictable resource usage, and programming abstractions that allow operators to specify compression policies safely.

\textbf{Compression, privacy, and data governance.} 
Compressed feature logs can still reveal sensitive information about users and devices. 
Future work should investigate how compression can be co-designed with retention policies, aggregation, and anonymization so that reduced storage also translates into reduced exposure and clearer data governance.

These guidelines and challenges suggest that compression is no longer a low-level implementation detail, but a design knob that should be exposed and managed alongside sampling, aggregation, and model selection in modern monitoring systems.

%input{challenges}
\section{Conclusions}
This article examined task-aware lossy compression as a mechanism to enable scalable, feature-based network traffic analysis across both high-throughput core networks and resource-constrained IoT environments. By evaluating compression in terms of downstream analytics performance rather than signal reconstruction fidelity, we showed that substantial storage reductions can be achieved while maintaining stable task accuracy within practical operating regions.

Across scenarios, a consistent principle emerges: compression must be co-designed with the target analytics task and tuned based on task-level performance, not abstract distortion metrics. Treating compression as a first-class design dimension in network monitoring systems is essential to sustain scalability under growing traffic volumes, pervasive encryption, and heterogeneous deployment constraints.

%This article examined the role of task-aware lossy compression in enabling scalable, feature-based network traffic analysis from large core networks to resource-constrained IoT environments. By focusing on downstream analytics accuracy rather than on signal reconstruction fidelity, we showed that meaningful storage reductions can be achieved without compromising the reliability of network intelligence tasks.

%We highlighted how different operational constraints expose different compression levers through two representative use cases. In core networks, where feature logs are already compact and heavily optimized, feature-wise scalar quantization provides operationally relevant savings on top of lossless compression. In IoT settings, where long-term retention under strict resource constraints is a primary concern, combining feature selection with coarse quantization enables accurate device identification at very low storage rates per device.

%A common principle emerges across these scenarios: feature compression strategies should be designed and tuned with the target analytics task in mind, and evaluated in terms of task performance within practical operating regions rather than through abstract distortion metrics. Treating compression as a first-class design dimension in network monitoring systems can help future infrastructures remain scalable and efficient in the face of increasing traffic volumes, pervasive encryption, and diverse deployment environments.

\bibliography{refs}
\bibliographystyle{IEEEtran}

\section*{Biographies}

\begin{IEEEbiographynophoto}{Fabio Palmese}
received the M.Sc. degree in Computer Engineering and the Ph.D. degree in Information Technology in the telecommunications area from the Politecnico di Milano, Milan, Italy, in 2020 and 2025, respectively. From September 2023 to March 2024, he was a visiting student with the EEE Department, University College London, U.K. His research activities focus on network traffic analysis and Internet of Things topics. His Ph.D. research thesis focused on the IoT forensics, with specific attention to IoT network traffic collection and analysis for forensic investigations.
\end{IEEEbiographynophoto}

\begin{IEEEbiographynophoto}{Gabriele Merlach}
received the M.Sc. degree in Electronic and Computer Engineering from the University of Trieste, Italy, in March 2024. From June 2024 to June 2025, he was a research fellow at the same university, working on big data processing and passive network measurement. His research interests include the study of Internet protocols, big data analysis and large-scale traffic characterization.
\end{IEEEbiographynophoto}

\begin{IEEEbiographynophoto}{Damiano Ravalico}
completed his studies in October 2023 when he obtained his master's degree in Electronic and Computer Engineering from the University of Trieste, Italy. The following month he started his PhD at the same university. His main areas of interest are network measurements and cyber security.
\end{IEEEbiographynophoto}

\begin{IEEEbiographynophoto}{Martino Trevisan}
is an Associate Professor at the Department of Engineering and Architecture of the University of Trieste. He received his M.Sc. (2015) and Phd (2019) in Computer Engineering from Politecnico di Torino. During his career, he visited Télécom ParisTech (Paris), Cisco Systems labs (San Josè, US), AT\&T (Bedminster, US), and the Universidade Federal de Minas Gerais (Brazil). He has published more than 60 papers in prestigious journals and conferences in the field of networking and big data. His research interests are mainly focused on big data methodologies for Web and Internet analysis. He also studies the operation of Online Social Networks and their implications on user behaviour
\end{IEEEbiographynophoto}

\begin{IEEEbiographynophoto}{Alessandro E. C. Redondi}
Alessandro E. C. Redondi (Senior Member, IEEE) is Associate Professor with the Dipartimento di Elettronica, Informazione e Bioingegneria of the Politecnico di Milano, Italy. He received the MS in Computer Engineering in July 2009 and the Ph.D. in Information Engineering in February 2014, both from Politecnico di Milano. From September 2012 to April 2013 he was a visiting student at the EEE Department of the University College of London (UCL). In 2018 he was a vitising researcher at the Computer Network Research Group of Universitat Politecnica de Valencia. His research activities are focused on the design and optimization of wireless/IoT systems and on network data analytics.
\end{IEEEbiographynophoto}

\end{document}